\documentclass{iaus}
\usepackage{graphicx}
\usepackage{graphics}

\title[short title of paper] 
{A search for OH 6 GHz maser emission towards southern supernova remnants}

\author[McDonnell, Vaughan \& Wardle]   
{Korinne E. McDonnell, %
 Alan E. Vaughan \break \and Mark Wardle}

\affiliation{Department of Physics, Macquarie University,
Sydney, NSW 2109, Australia \break email: \{korinnem,
alanv, wardle\}@physics.mq.edu.au \\
}

\pubyear{2007}
\volume{242}  
\pagerange{119--126}
\date{?? and in revised form ??}
\jname{Astrophysical Masers and their Environments IAU Symposium}
\editors{J.M. Chapman \& W.A. Baan, eds.}
\begin{document}

\maketitle

\begin{abstract}
OH masers at 1720~MHz have proven to be excellent indicators of
interactions between supernova remnants and molecular clouds.  Recent
calculations suggest that the 6049~MHz OH maser line is excited for
higher column densities than for the 1720~MHz line.  It is therefore a
potentially valuable indicator of remnant-cloud interaction.

We present preliminary results of a survey using the Parkes Methanol Multibeam receiver for 6049 MHz
and 6035/6030~MHz OH masers towards 36 supernova remnants and 4 fields
in the Large and Small Magellanic Clouds.  While no 6049~MHz masers
have been found, three new sites of 6035 and 6030~MHz OH maser
emission have been discovered in star-forming regions.
\end{abstract}


OH masers at 1720~MHz have proven to be an almost unambiguous
indicator of the interaction between supernova remnants and molecular
clouds (\cite{Frailetal1994}).  OH excitation calculations (\cite{PavlakisKylafis};
Wardle, these proceedings) suggest
that the 6049~MHz satellite line may be present at higher OH column densities where the 1720~MHz
line is weak or absent.


A survey using the Parkes Methanol Multibeam
  receiver at 6049 and 6035/6030~MHz was conducted towards 36
  supernova remnants and 4 fields in the Large and Small Magellanic
  Clouds.  An 8~MHz bandwidth with 2048 channels was used, allowing a
  channel width of 0.2~km~s$^{-1}$.  The beam size is approximately 3.3
  arcminutes and both circular polarisations were observed.  Scans in right ascension and declination were conducted
over the supernova remnants, with a total observing time in one direction of $\sim$100
minutes per square degree.  The data was flux-calibrated, continuum subtracted and gridded using
the programs Livedata and Gridzilla.  The data cubes produced were
searched for maser emission using the program Duchamp (\cite{Whiting}).

\begin{figure}
\begin{center}
 \includegraphics[scale=0.87]{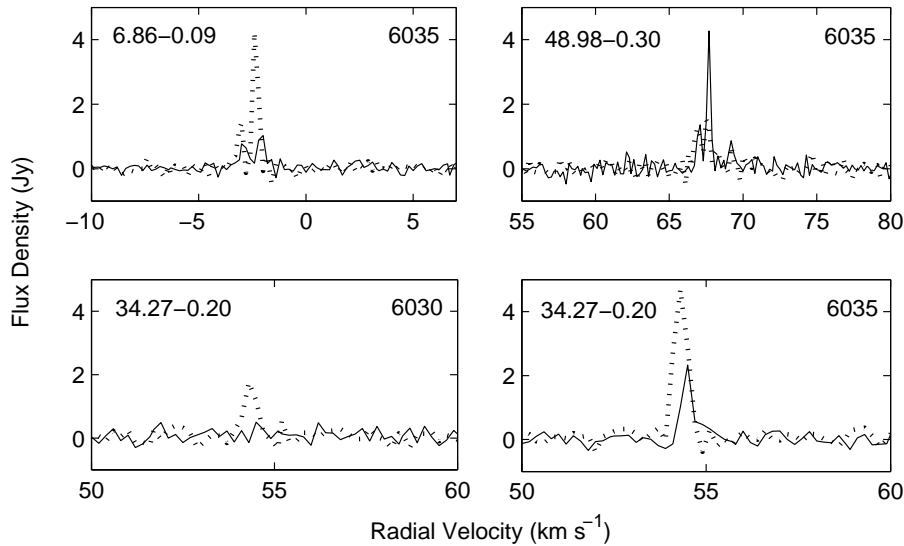}
\end{center}
  \caption{Spectra of new OH masers at the 6030- and 6035-MHz
    transitions.  LHCP and RHCP are displayed as dashed and solid
    lines, respectively.  The channel resolution is 0.2~km~s$^{-1}$
    and the beamsize is approximately 3 arcminutes.}\label{fig:4figures}
\end{figure}

While preliminary analysis has not discovered any 6049~MHz maser
emission, 5 maser sites at 6035~MHz have been identified (2 containing 6030~MHz
emission as well).  Masers 6.86-0.09, 34.27-0.20 and 48.98-0.30 are new
discoveries (see Fig. \ref{fig:4figures}), while 336.941-0.156 and 337.705-0.053 are already
known (\cite{CaswellVaile}; \cite{Caswell}).

As left and right-hand circular polarisations (LHCP and RHCP) were observed, Zeeman
pairs can be recognised.  A 1~mG magnetic field produces splittings
equivalent to 0.079~km~s$^{-1}$ and 0.056~km~s$^{-1}$
in the 6030 and 6035~MHz transitions respectively.  The magnetic fields calculated can be
found in Table \ref{Tab:masers} and have an uncertainty of
approximately 2~mG.

The 6035~MHz maser discovered at 48.98-0.30 is approximately
coincident with two H~\small{\textsc{II}} \normalsize regions and is likely to be associated with
one of them.  It is expected that the other 6030 and 6035~MHz masers
are associated with star forming regions. 

These results are preliminary and further analysis may yield weaker 6~GHz masers.  In addition, data from ATCA observations of seventeen supernova remnants remain to be analysed.        
\begin{table}
\begin{center}
\caption{Masers at the 6035- and 6030- MHz OH transitions.  The
  velocity and flux of the peaks are taken at the brightest peak, if
  multiple peaks are observed.}
\label{Tab:masers}
\begin{tabular}{lllllll}
\hline
	& 	&              &	6035 MHz &	6035
        MHz & & 6030 MHz \\
OH maser&RA (2000)&Dec (2000)	& Velocity Peak  & Peak flux &
Magnetic & Peak  \\
(l b)	&	&             &LHCP\,\,\,\, RHCP & LHCP\,\,\,\,RHCP &
Field & (L and/or R)	\\
($^{\circ}$ $^{\circ}$)	&(h m s)&   ($^{\circ}$ $'$ $''$)   &(km
s$^{-1}$) & (Jy) & (mG) & (Jy)	\\
\hline							
336.941-0.156 & 16 35 55.20&-47 38 45.4	&-65.6\,\,\,\,\, -65.1
& 3.35\,\,\,\, 1.86 & +8.9 & 1.03\,\,\,\, 0.46 \\
337.705-0.053 & 16 38 29.67	&-47 00 35.8	&-53.6\,\,\,\,\, -50.7&
1.63\,\,\,\, 2.22 & \,\,\,\,- & \,\,\,\,- \\
6.86-0.09 &	18 00 48	&-22 58 14	&-2.37\,\,\,\,\, -1.98
&4.23\,\,\,\, 1.03  & +7.0 & \,\,\,\,-	\\
34.27-0.20	&18 54 36	&+01 05 54	&\,54.3\,\,\,\,\,\,\,\, 54.5
&4.60\,\,\,\, 2.33 & +3.6 &1.76 (L)	\\
48.98-0.30& 19 22 27	&+14 06 53	& \,67.5\,\,\,\,\,\,\,\, 67.7
&1.55\,\,\,\, 4.27    &	+3.6 & \,\,\,\,- \\
\hline
\end{tabular}
\end{center}
\end{table}

\begin{acknowledgments}
We would like to thank Catherine Braiding for helping with the
observing, Stacy Mader for observing assistance and help with data reduction, and James Caswell for
assistance with maser identification.
\end{acknowledgments}

\end{document}